# Absolute measurement of penetration depth of superconducting thin films using microwave stripline resonators

Arghya Dutta[1*], Ajeet Salunke[1*], Mahesh Poojary[2], Vivas Bagwe[1], Sangita Bose[2] and Pratap Raychaudhuri[1†]

[1] *Tata Institute of Fundamental Research, Homi Bhabha Rd, Mumbai 400005, India*

[2] *School of Physical Sciences, UM-DAE Centre for Excellence in Basic Sciences, University of Mumbai, Kalina Campus, Santacruz (E), Mumbai 400098, India*

Superconducting microstrip resonators, which leverage kinetic inductance to probe electrodynamics, are sensitive tools for studying superconducting thin films at microwave frequencies. However, extracting the absolute superconducting penetration depth, $\lambda$, from these measurements remains challenging. In this work, we present a hybrid method to determine the absolute value of $\lambda$ over a wide temperature range by combining resonator measurements with finite-element electromagnetic simulations in COMSOL Multiphysics. We demonstrate this approach by extracting the penetration depth of NbN and $Nb_3Sn$ films by fabricating resonators from films of various thicknesses. Furthermore, we extend the technique to materials with lower critical temperatures by employing a flip-film geometry. By placing a sample above a NbN resonator, separated by a thin Mylar dielectric, we create a coupled structure where changes in the sample's penetration depth shift the resonant frequency. This non-destructive method provides a reliable, high-sensitivity platform for characterizing the penetration depth of diverse superconducting thin films.

[*] These authors contributed equally.
[†] E-mail: pratap@tifr.res.in

## I. Introduction

The magnetic penetration depth, $\lambda$, is a key parameter governing the electrodynamic response of superconductors and provides important insight into the superconducting state[1,2]. Among different techniques used to measure $\lambda$, resonant techniques operating from tens of megahertz to gigahertz[3,4,5] where the superconductor forms part of the resonant cavity, are by far the most sensitive to small changes in penetration depth. The operating principle relies on the fact that the inductance of the cavity depends on the penetration depth of the superconductor. Thus, small changes in $\lambda$ results in changes in the resonant frequency, which can be used to measure the change in $\lambda$ with temperature. However, in all these techniques, determining the absolute value of $\lambda$ by accounting for all geometrical factors remains a major challenge. Consequently, the absolute value of $\lambda(0)$ in the zero-temperature limit is often estimated from other measurements: normal state resistivity and superconducting energy gap[5], or from the lower and upper critical fields[6,7,8], $H_{c1}$ and $H_{c2}$, or by assuming a specific functional form for the temperature dependence[9,10,11,12] of $\lambda$. However, these methods are intrinsically model dependent and work best with conventional superconductors that follow Bardeen-Cooper-Schrieffer (BCS) theory. Another method involves a complex calibration by comparing the a.c. conductivity with the d.c. conductivity[13,14,15,16] above the superconducting transition temperature. However, this method cannot be applied in all situations, for example, in superconducting planar resonators where the resonance characteristics in the normal state often become too broad to resolve any resonance peak. Alternatively, $\lambda(0)$ can be determined using other non-resonant techniques, such as low-frequency planar two-coil mutual inductance technique[17,18,19], or broadband microwave corbino spectroscopy[20,21,22,23,24,25] or muon spin rotation measurements[26,27]. Of these, the planar two-coil mutual inductance technique has a sensitivity comparable to resonant techniques for thin films. So far, one resonant technique using variable spacing parallel plate resonator has been developed[28,29] to obtain the absolute value of superconducting penetration depth at microwave frequencies. However, the experimental setup is mechanically complex, involves fine positioning of the sample and has not been widely used.

In this paper, we present a new method to determine the absolute value of $\lambda$ in superconducting thin films over a wide range of temperatures by combining measurements on superconducting microstrip resonators with numerical solution of the Maxwell equations using COMSOL Multiphysics® simulation package[30]. Superconducting microstrip resonators, where a microwave cavity is created by patterning a narrow stripline of finite width, $w$, and length, $l \gg w$ on a superconducting thin film, are widely used as a versatile device both for basic science and applications in high frequency superconducting electronic devices[31,32,33,34]. The resonant frequency of the stripline resonator[35] is determined by the wavelength of the standing wave supported in the stripline and the phase velocity of electromagnetic wave, $v_{ph}$. For an open stripline in ideal condition, the length of the stripline, $l$, equals to half the wavelength at the fundamental resonant frequency, $f_1$, and the resonant frequencies are given by[36] $f_n = n\frac{v_{ph}}{2l} = \frac{n}{2l}\frac{c}{\sqrt{\varepsilon_{eff}}}$ $(n = 1,2, \ldots)$, where $c$ is the speed of light in vacuum and $\varepsilon_{eff}$ is the effective dielectric constant. This can be expressed in terms of the inductance, $L$ and capacitance, $C$ per unit length of the microstrip as $f_n = \frac{n}{2l}\frac{1}{\sqrt{LC}}$ $(n = 1,2, \ldots)$. When the stripline is made out of a normal metal, $L$ and $C$ are given by the geometric inductance and capacitance, $L_G = \frac{Z_0\sqrt{\varepsilon_{eff}}}{c}$ and $C_G = \frac{\sqrt{\varepsilon_{eff}}}{Z_0 c}$, where $Z_0$ is the characteristic impedance of the microstrip. When the stripline is in the superconducting state, another component, the kinetic inductance[37] (per unit length), $L_k = g\frac{\mu_0\lambda^2}{A}$, arises from the response of the superfluid and gets added to the geometric inductance; $\mu_0$ is the vacuum permeability, $A$ is the cross-sectional area of the stripline and $g$ is a geometric factor that takes into account non-uniformities of current distribution in a stripline of a given geometry. Consequently, the modified resonant frequency of the superconducting microstrip, $f_n^s = \frac{n}{2l}\frac{1}{\sqrt{C(L_G+L_k)}}$, now encodes information on the penetration depth. However, several practical difficulties limit the determination of $\lambda$. First, it is difficult to accurately determine $Z_0$ and $g$, and approximate analytic expressions and empirical formulae based on numerical calculations exist only for certain geometries[38,39,40,41]. Secondly, the real microstrip is never completely isolated from its surroundings. It is coupled to the ground planes, input launching pads and the grounded surfaces of the box, all of which influence the resonant frequencies. Furthermore, space considerations in a cryogenic experiment often necessitates the use of meander microstrip structure instead of a simple straight microstrip, further complicating the electric and magnetic field distribution. Consequently, the resonant frequency shifts from its ideal value.

To circumvent these problems, we adopt a "brute force" approach. We accurately model the experimental configuration including details of the device and the sample box and numerically solve the Maxwell equations to find the resonant frequencies. For superconducting surfaces, we include an imaginary part to the complex conductivity ($\sigma'' = 1/(\mu_0\omega\lambda^2)$) and adjust $\lambda$ to precisely match the resonant frequencies. In the first part of this work, we present results on microstrip resonators made of NbN and $Nb_3Sn$ and show that this technique provides absolute values of $\lambda$ for films with thickness varying from 6-100 nm. In the second part of this work, we show that the technique can be extended to measure the penetration of another continuous superconducting film, by placing the film upside-down with a thin Mylar spacer layer on the top a well characterized NbN microstripline, in a typical a flip-film geometry. We demonstrate this technique by measuring the penetration of a 6 nm thick NbN film and an amorphous $Re_6Zr$ ($a$-$Re_6Zr$)[42]. Analyzing the resonant frequency of this coupled structure we can determine the $\lambda$ of the unpatterned film. Our technique establishes microstrip resonators as a versatile quantitative tool to study the penetration depth of superconductors.

## II. Experimental Details

The superconducting microstrip resonators are fabricated on superconducting NbN films of different thicknesses and a 90 nm thick $Nb_3Sn$ thin film, grown on single crystalline [100] oriented MgO substrates using d.c. magnetron sputtering. Details of thin film growth and characterization are reported elsewhere[43,44]. The microstrip is patterned on the films using a maskless photolithography process using an ultraviolet laser writing system (Holmarc, HO-LWS-PUV-T) followed by argon ion beam milling. Fig. 1(a)-(b) show the geometry of the device. The microstrip consists of a meander with a width of 13-19 μm and a length of 19.23 mm. The variation in the resonator widths is mainly due to variations in a lithographic process. However, all device dimensions were independently measured under a microscope, and the measured dimensions were used during simulations. The microstrip is enclosed in a $2.4\,mm\ \times 2.6\,mm$ box patterned on the ground pad. We confirmed from experiments and simulations that this configuration increases the quality factor of the resonator, while still allowing sufficient coupling to a film placed above the resonator in the flip-film geometry (Fig. 1(d)). The microstrips are capacitively coupled to the microwave launching pads at the two ends, with capacitive gap of 60 μm, corresponding to a coupling capacitance of about 3.5 fF. The device is enclosed in gold-plated copper box fitted with two SMA connectors as shown in Fig. 1(c). To measure λ in the flip-film geometry, we use two different continuous films grown on 5 mm × 1.9 mm MgO substrate: (i) a 28 nm thick amorphous $Re_6Zr$ film ( $T_c$~5.67 K ) grown using pulsed laser deposition[42] and (ii) a 6 nm thick NbN film grown through reactive magnetron sputtering. We place these films upside down just above a 60 nm thick NbN microstrip resonator with a 25 μm thick mylar spacer in between, as shown in Fig. 1(d). The surface of the film is grounded from the edge with a thin gold wire. A spring-loaded Teflon cylinder is used to hold the sample on the resonator in place.

The microwave characteristics and resonant frequency of the microstrip were measured in the transmission mode. We measured both the amplitude and phase of the transmission coefficient, $S_{12}$, using a vector network analyzer (Rhode & Schwarz, ZVB). Low temperature measurements are done in a continuous flow $^4$He cryostat operating down to 2 K. To reduce heat load, the microwave transmission lines were made from a combination of rigid stainless steel coaxial cables connecting from room temperature to the low temperature stage, followed by flexible copper coaxial cables. The input microwave line is fitted with a 40 dB attenuator. At the lowest temperature (2.1 K) we keep the output power from the vector network analyzer at -30 dBm, which corresponds to ~ -70 dBm after the attenuator. With an increase in temperature the resonance peaks typically become broader, and we gradually increase the power but always keep it below -40 dBm on the sample.

## III. Details of Simulation

To determine the penetration depth, the general strategy is as follows. For a given geometry, we use COMSOL Multiphysics Software to compute the resonant frequency of the microstrip by solving the Maxwell equations using finite element analysis. We construct a look-up table by computing the resonant frequencies $f_n$ for different values of λ. We then find the value of λ corresponding to the measured $f_n$ by interpolating this curve. The accuracy of this method is crucially dependent on the precise modelling of the experimental configuration which includes the actual device as well as the box inside which the sample is loaded. We model the entire experimental configuration as shown in Fig. 1(c) or Fig. 1(d). The gold-plated copper surfaces of the box are assumed to be perfect conductors. The insulating substrate and spacer layer were defined by their respective relative dielectric constants, $\varepsilon_r$. The values of $\varepsilon_r$ for both MgO substrate and Mylar spacer are important parameters in our simulation. For MgO substrates, $\varepsilon_r$ at low temperature has been precisely measured at 10.5 GHz and varies from[45] 9.5-9.7. From simulations, we find that this variation results in an uncertainty of $\sim \pm 4\%$ in the value of $\lambda$. Therefore, we fix this value to $\varepsilon_r(\mathrm{MgO}) = 9.6$. For Mylar, reported values of $\varepsilon_r$ at room temperature vary in the range[46,47,48,49] 2.8-2.48 in the frequency range 1 GHz-100 GHz. However, we did not find any report at low temperatures. Therefore, we independently determined this value as $\varepsilon_r(\mathrm{Mylar}) = 2.48$, by placing a mylar sheet on the top of a superconducting microstrip resonator and measuring the resulting shift in the resonant frequency at 2.4 K and then comparing with simulations. This value was used in the final simulations.

To simulate the superconducting thin films, we use a transition boundary condition where the discontinuities in the tangential electric and magnetic field are related to the film's conductivity, $\sigma$, and thickness, $t$, without explicitly creating a mesh inside the volume of the superconductor[50]. This method is computationally efficient and very reliable for conductive layers. We use the fact that for a superconductor at frequencies well below the superconducting gap the conductivity $\sigma_s$ has both a real and imaginary part, given by, $\sigma_s = \sigma' - i\sigma''$, where $\sigma'$ is the dissipative component and $\sigma''$ is the inductive component. Within two-fluid approximation[51], the charge carrier density, $n$, can be separated into a superfluid component, $n_s\left(=\frac{m}{\mu_0 e^2\lambda^2}\right)$, and a normal fluid component $n_n = n - n_s$, such that $\sigma' \approx \left(\frac{\pi n_s e^2}{m}\right)\delta(\omega) + \frac{n_n e^2\tau}{m}$ ($\tau$ is the normal electron scattering time) and $\sigma'' = \frac{n_s e^2}{m\omega} = \frac{1}{\mu_0\omega\lambda^2}$. The resonant frequency of the microstrip is almost entirely determined by $\sigma''$ which is the quantity that depends on $\lambda$; $\sigma'$ determines the quality factor of the resonator but has almost no influence on the resonant frequency as long as $\sigma'/\sigma'' \ll 1$ (see Appendix 1).

This condition is valid in the superconducting state except very close to $T_c$. Therefore, in our simulations we keep a fixed value, $\sigma' = 10^3 - 10^4\ \mathrm{S\cdot m^{-1}}$, for which this condition is satisfied over the entire range of our measurements.

An important factor in our analysis is the mesh used for finite element analysis. A sparse mesh introduces large errors, whereas a dense mesh increases the computation time. Since the electric and magnetic variation is largest in the vicinity of the microstrip, we use a very dense mesh on the microstrip with maximum element size of 5 μm. Furthermore, at the edges of the ground plane and on the microwave launching pads we use a medium mesh density with maximum element size of 50 μm. Elsewhere on the substrate and in the box, we use the default mesh in COMSOL which has a maximum element size of 4.5 mm. The internal routine of COMSOL optimizes the mesh to ensure smooth transition from one mesh size to another. With this mesh we can compute the resonant frequency for a given value of λ in 10-15 minutes on a mid-level gaming computer ($12^{th}$ generation i-9 processor and 64 Gb RAM). When we reduce the dense mesh maximum element size to 4 μm the computation time increases roughly by 4 times, while the resonant frequency changes by less than 0.04%. On the other hand, reducing the maximum element size of the medium and default mesh even by a factor of 10 does not change the resonant frequency by more than 0.03%. The final mesh used in our simulations is shown in Fig. 1(e) and 1(f).

## IV. Experimental Results

We first concentrate on a series of microstrip resonators made from NbN thin films with different thicknesses. Fig. 2(a) shows the $|S_{12}|$ as a function of frequency between 500 MHz - 16 GHz at 2.2 K for the resonator fabricated on a 60 nm thick NbN film with $T_c \sim 14.8$ K. The resonances of the microstrip manifest as sharp peaks in $S_{12}$. In addition, between 9-11 GHz and close to 16 GHz we observe some broad peaks, associated with parasitic resonance from the box. These broad peaks are temperature independent and not associated with the superconducting film. We can clearly resolve the 4 peaks corresponding to the fundamental frequency $f_1$ and the higher harmonics $f_2, f_4$ and $f_5$. $f_3$ is masked between two parasitic peaks and cannot be unambiguously resolved. Fig. 2(b) shows an expanded view of the fundamental resonance peak obtained from a higher resolution frequency scan. The unwrapped phase, shown in the same plot undergoes an $180^0$ phase shift across the resonance, as expected. To accurately determine the resonant frequency, we fit the lineshape of $S_{12}$ with the theoretical expression,

$$|S_{12}| \propto \left(\frac{1}{1+4Q_L^2\delta^2}\right)^{1/2}, \qquad (1)$$

where $Q_L$ is the loaded q-factor and $\delta = \frac{f - f_n}{f_n}$. From the fit we obtain $f_1 = 4.08515$ GHz and $Q_L = 3534$. The inset of Fig. 2(b) shows the computed value of $f_1$ for different values of λ for NbN. The experimental value of $f_1$ corresponds to $\lambda = 276$ nm. This is remarkably consistent with the reported value[18] $\lambda = 275$ nm for a NbN film of similar thickness (50 nm) obtained from low frequency mutual-inductance technique. The largest source of errors in the values of λ in our measurement is from the uncertainty in the film thickness (typically $\pm$ 4%) and to a smaller extent from the substrate-to-substrate variation[45] in the value of $\varepsilon_r$(MgO). Combining these two we estimate a maximum uncertainty of $\pm$8% in the value of λ. To further verify the reliability of our simulation, we also calculate the higher harmonic frequencies $f_2 - f_5$ using the same value of λ. As can be seen in Fig. 2(c), the experimental values of $f_2, f_4$ and $f_5$ match very closely with the computed values. We observe that $f_n$ $vs. n$ deviates from a linear relation. This non-linearity has been attributed to long range Coulomb interactions with the ground plane below the device[52]. It is remarkable that this nonlinearity is fully accounted for in our simulations. To investigate the temperature dependence of λ, we track $f_1$ from the evolution of $|S_{12}|$ as a function of temperature (inset of Fig. 2(d)). With increase in temperature $f_1$ shifts to lower frequencies, consistent with an increase in λ, and hence, $L_k$. At the same time the resonance peak gradually becomes broader, and we can resolve the resonance peak up to $0.96T_c$. The temperature dependence of $f_1$ is shown in the *inset* of Fig. 2(d). Adopting the same procedure as before, we determine λ at each temperature. In Fig. 2(e), we plot the temperature dependence of $\frac{1}{\lambda^2}$ ($\sim n_s$) (the *inset* shows the temperature variation of $\lambda$). For a conventional superconductor in the dirty-limit, this variation is given by the dirty-limit BCS expression[51],

$$\frac{1}{\lambda(T)^2} = \frac{1}{\lambda(0)^2}\frac{\Delta(T)}{\Delta(0)}\tanh\left[\frac{\Delta(T)}{2k_B T}\right], \qquad (2)$$

where $k_B$ is the Boltzmann constant and Δ is the superconducting energy gap. We obtain very good fit of the experimental data with eqn (1) assuming the BCS temperature variation[51] for Δ and using $\Delta(0)$ and $\lambda(0)$ as fitting parameters. The best fit value of $\lambda(0)$ is practically the same as the value at 2.2 K; on the other hand, $\Delta(0) = 2.88$ meV is consistent with the values obtained from tunneling measurements[18,53,54] on NbN films with similar $T_c$. Fig. 2(f) show $\frac{1}{\lambda^2}$ $vs. T$ obtained from resonators made of different NbN thickness. $\frac{1}{\lambda(0)^2}$ decreases with decreasing film thickness as reported earlier both for NbN[18,55] and other superconductors[56]. For the resonators made of 6 nm thick NbN film, the resonance peaks gets broad faster with increasing temperature, yet we can resolve $f_1$ up to $0.89T_c$ respectively. Fig. 2(g) and (h) show $|S_{12}|$ vs. $f$ and $\frac{1}{\lambda^2}$ $vs. T$ for a similar microstrip resonator made from 90 nm thick $Nb_3Sn$ film. It is worthwhile to note that the extracted values of $\lambda(0)$ for NbN are close to the values obtained from low-frequency (30 kHz) mutual inductance technique[18], except for the 6 nm thick film for which $\lambda(0)$ is higher by a factor of 1.9. This may be caused by vortex-antivortex pairs in ultrathin superconducting films, which appear when film thickness is near or below the coherence length[57]. Close to $T_c$, the unbinding of these pairs leads to the Berezinskii-Kosterlitz-Thouless (BKT) transition, at which the superfluid density abruptly drops to zero[55,58,59]. On the other hand, below the BKT transition temperature, these pairs are tightly bound and do not contribute to d.c. transport or the superfluid response at low frequencies. However, at microwave frequencies the antiphase motion of the vortex and antivortex in the microwave magnetic field can produce an additional component to the kinetic inductance, which would result in an effective increase in λ. At the same time 2-D quantum confinement has also been predicted to increase $\lambda(0)$ in ultrathin films, though the frequency dependence of this effect has not yet been explored[60]. Therefore, the precise origin of the enhanced value of $\lambda(0)$ in very thin films will need to be confirmed from further investigations.

We now discuss the measurement of penetration depth in the flip-film geometry. We first concentrate on the measurement on a 28 nm thick continuous *a*-$Re_6Zr$ film. We place the *a*-$Re_6Zr$ film on a microstrip resonator made from 60 nm thick NbN films as shown in Fig. 1(d). Fig. 3(a) shows the progressive shift of the fundamental resonance peak at 2.1 K (i) for the bare microstrip, (ii) when a 25 μm Mylar spacer is placed on the top and (iii) when *a*-$Re_6Zr$ film is placed on the top of Mylar spacer. First, we analyze configuration (ii), where only the Mylar spacer is placed on top of the microstrip, to determine $\varepsilon_r$ for the Mylar spacer layer. We calculate $f_1$ for this configuration as a function of $\varepsilon_r$ of the spacer layer (Fig. 3(b)). The penetration depth for microstrip is fixed to, $\lambda_{NbN}^{\mu\text{strip}} = 276$ nm, determined from the resonant frequency of the bare resonator. By matching with the experimental resonant frequency, we obtain $\varepsilon_r(Mylar) = 2.48$. In the next step, we construct a look up table by calculating $f_1$ for configuration (iii), where we vary the value of λ for the film placed on the top, while keeping $\lambda_{NbN}^{\mu\text{strip}} = 276$ nm, $\varepsilon_r(Mylar) = 2.48$ constant (Fig. 3(b)). Comparing the experimental resonant frequencies at various temperatures (shown in Fig. 3(c)) with this simulated curve, we determine the temperature variation of λ for the *a*-$Re_6Zr$ film (Fig. 3(d)). The temperature variation of $\frac{1}{\lambda^2}$ closely follows equation (2) with $\lambda(0) = 1039\ nm$ and $\Delta(0) = 0.96\ meV$. Both values are consistent with the values reported from two-coil mutual inductance and tunneling respectively[42]. Since the measurement gets affected by both the thickness error of the microstrip resonator and *a*-$Re_6Zr$ film, we estimate an error of $\pm 12\%$ in the absolute value when the penetration depth is measured with this method. In this analysis, we have neglected the temperature variation of $\lambda_{NbN}^{\mu\text{strip}}$, which is ~ $0.5\%$ over the entire range of measurements. This introduces a maximum error of $2\%$ in $\lambda$ for the *a*-$Re_6Zr$ film for temperatures close to $T_c$, which does not alter the temperature dependence in any significant way.

As a second example of measurement using the flip-film geometry we measured the penetration depth of a continuous 6 nm NbN film ($T_c \sim 11.2\ K$) using the 60 nm NbN microstrip resonator. We can resolve the resonance for $f_1$ from 2 K to 10 K (Fig. 4(a)). Here, since the $T_c$ of the film and the microstrip are close to each other we cannot ignore the temperature variation of $\lambda_{NbN}^{\mu\text{strip}}$ (276 nm – 350 nm over the temperature range of measurement). To account for this, we construct a two-dimensional look up table for $f_1$ by varying both $\lambda_{NbN}^{\mu\text{strip}}$ and the penetration depth of the sample placed in the flip film configuration, $\lambda_{\text{sample}}$. The $\lambda$ value for 6 nm NbN films is extracted at each temperature by matching experimental $f_1$ values to the lookup table data using the corresponding $\lambda_{NbN}^{\mu\text{strip}}$ at that temperature (Fig. 4(b)). Fig. 4(c) shows the temperature variation of $\lambda$ and $\frac{1}{\lambda^2}$ of the 6 nm NbN film and the corresponding fit with eqn (2).

The main advantage of the flip-film method is that it avoids the complicated process of patterning the sample in a microstrip resonator. However, there is one limitation. Here, the superconducting film is weakly coupled to the microstrip resonator through the Mylar spacer layer. In this configuration, the geometric effect is large and the change in $f_1$ due to $L_k(\propto \lambda^2/t)$ of the sample is generally much smaller than when the measurement is performed directly from the resonator characteristics. This poses problems for thick films where geometric effects dominate over $L_k$. Even though the measurements continue to give a reproducible shift of $f_1$ with temperature (down to resolution ~ 0.001 GHz), our numerical simulation loses the accuracy to reliably determine this small frequency change from which the absolute value of $\lambda$ can be determined (see Appendix 2). From measurements on many samples, we observe that this technique works well only with films that are thinner than $\lambda/4$. Nevertheless, since many applications in superconducting microwave electronics (such as hot electron bolometers) rely on very thin films that have large penetration depth, this method provides a quick and convenient way to characterize those samples.

## V. Summary

We presented a versatile technique to measure the absolute value of the superconducting penetration depth using superconducting microstrip resonators by combining precise measurements of the resonant frequency and numerical simulations. We have demonstrated two variants of this technique. In the first variant, we measure the resonant frequencies as a function of temperature of a microstrip resonator fabricated using the material under investigation. The second variant is based on using a superconducting microstrip resonator with known characteristics, to measure the penetration depth of another superconducting film with lower $T_c$ by coupling it to the resonator. This work removes a long-standing bottleneck, namely, determining the absolute value of superconducting penetration depth using microwave resonator technique. In principle, this technique can also determine a superconducting film's surface resistance, $R_s$, using the cavity's Q-factor. However, our experience is that accurately computing the Q-factor of the superconducting microstrip is numerically and computationally far more demanding. Future studies will concentrate on improving numerical methods to accurately simulate the Q-factor of the resonator structures from which $R_s$ can be determined.

*Acknowledgements:* The authors would like to thank Marc Scheffler and Rajamani Vijayraghavan for valuable discussions, Steve Anlage and Danile Torsello for valuable feedback on the manuscript and John Jesudasan for technical help. This work was supported by the Department of Atomic Energy, Government of India. AD, MP, SB synthesized the thin films. AD fabricated the microstrip resonators. AD and AS performed microwave measurements and analyzed the data. VB provided technical support for the experiments. PR and SB conceptualized the project and wrote the paper with inputs from all authors.

*Data availability:* All data used in this paper and COMSOL simulation files are available from the corresponding author upon reasonable request.

## Appendix 1: Influence of $\sigma'$ on the resonant frequencies

In our simulations we have kept the $\sigma'$ of the superconductor constant and only varied $\sigma''$ when calculating the resonant frequency of our devices. In fig. 5(a) we plot the calculated $|S_{12}|$ corresponding to the fundamental resonance, for different values of $\sigma'$ using $\sigma''$ and device parameters corresponding to the 60 nm thick NbN microstrip resonator shown in Fig. 1(e). We observe that effect of increasing $\sigma'$ is to broaden the resonance peak, but the resonance frequency (Fig. 5(b)) remains practically unaltered as long as $\frac{\sigma'}{\sigma''} < 0.01$. This condition is satisfied for most superconductors[4] except very close to $T_c$. In our simulations for the NbN resonators we keep $\sigma'$=10000 which is much smaller than $\sigma''$ over the entire range of measurement (*inset* Fig. 5(b)).

## Appendix 2: Measurement Sensitivity in the flip-film geometry

Fig. 6 shows the calculated values of $f_1$ for a microstrip resonator with a superconducting film mounted in the flip-film geometry as a function of the penetration depth of the film, $\lambda_{\text{sample}}$. We observe that the sensitivity of $f_1$ to $\lambda_{\text{sample}}$ progressively decreases as the film thickness increases. The thick limit in the figure corresponds to a semi-infinite superconductor placed on the microstrip resonator.

**References:**

[1] R. Prozorov and R. W. Giannetta, Magnetic penetration depth in unconventional superconductors. *Supercond. Sci. Technol.* **19** R41 (2006).

[2] P. Raychaudhuri and S. Dutta, Phase fluctuations in conventional superconductors. *J. Phys.: Condens. Matter* **34** 083001 (2022).

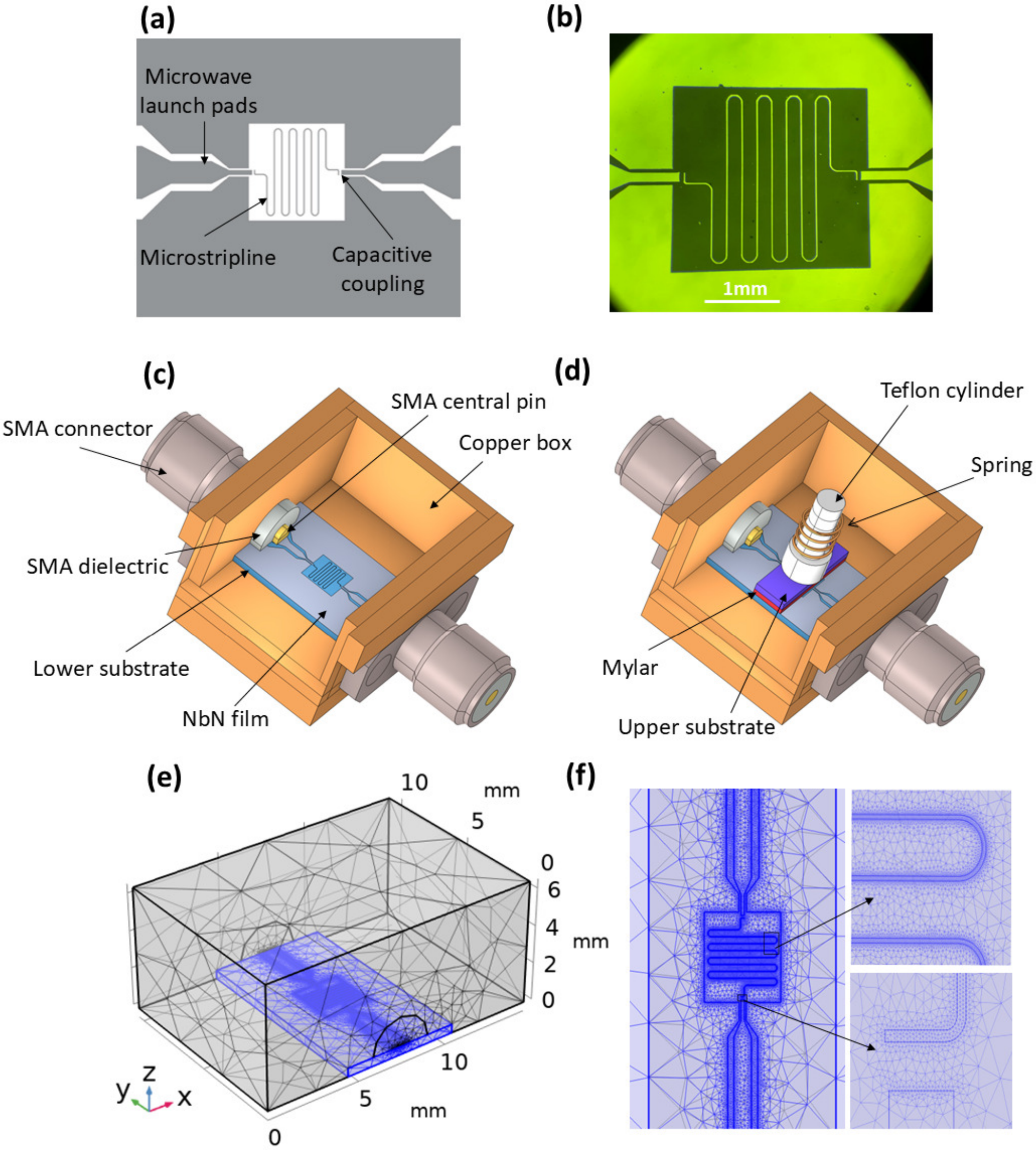

**Figure 1.** (a) Schematic of the microstrip resonator device and (b) optical image of a device fabricated from a 60 nm thick NbN film. The meander is located inside a rectangular box patterned on the ground plane. Experimental configuration (not to scale) of (c) a bare NbN microstrip resonator mounted in a copper box and (d) where a second superconducting film is mounted upside-down on NbN microstrip resonator with a 25 µm Mylar spacer in between in flip-film geometry; the central pin of the SMA connector is soldered on the launch pad of the device. The top and front surfaces of the box are hidden. (e) The mesh used for electromagnetic simulations; (f) expanded view of the mesh close to the microstrip and launching pads.

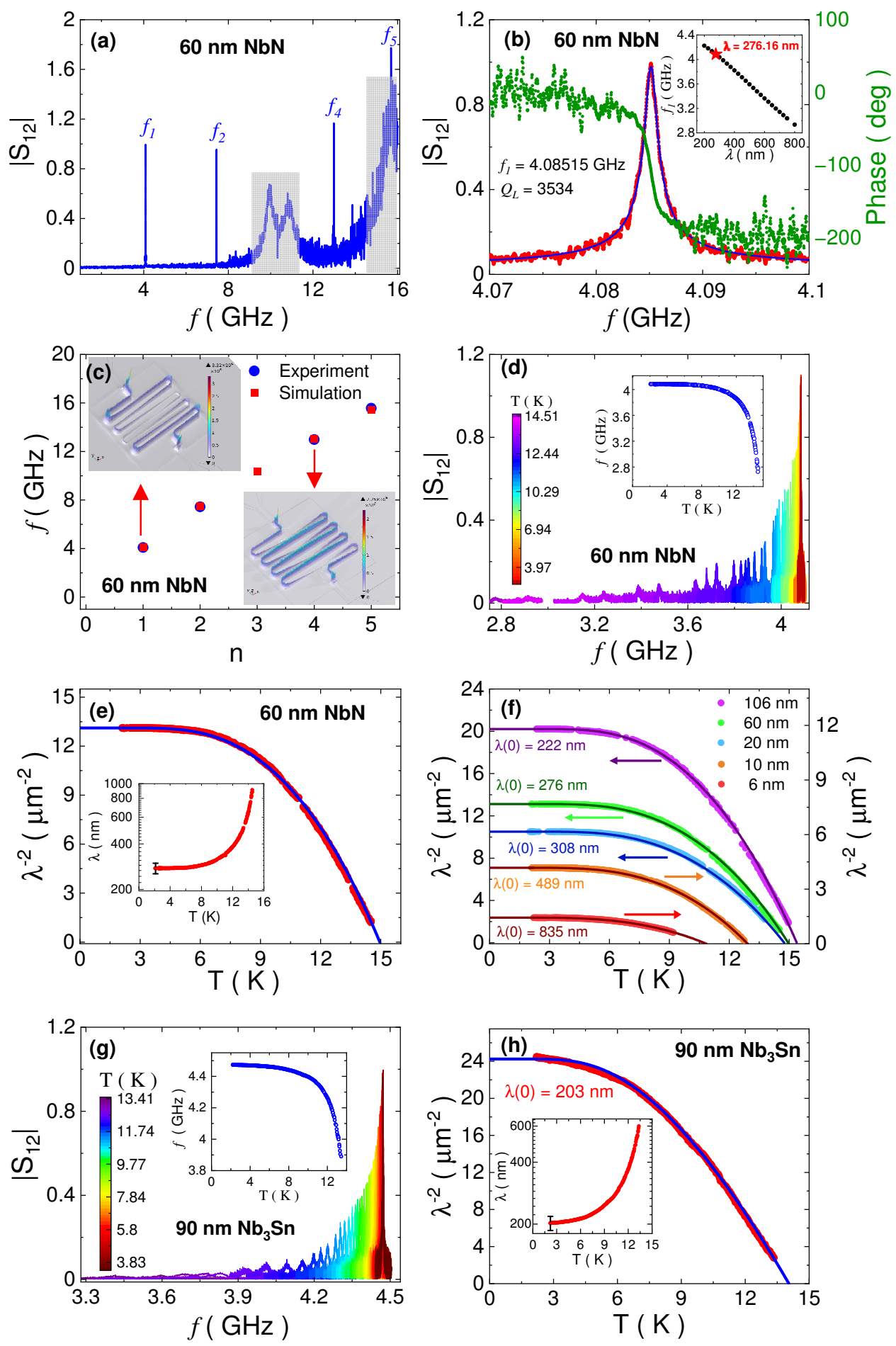

**Figure 2.** (a) Measured transmission coefficient $|S_{12}|$ as a function of frequency for a microstrip resonator made from 60 nm thick NbN film at 2.2 K. The sharp peaks labelled $f_1, f_2, f_4$ and $f_5$ are the resonant frequencies of the microstrip. The broad shaded features are parasitic resonances from the sample box. (b) Expanded view of $|S_{12}|$ and the phase across the fundamental resonance peak $f_1$; the solid line is a fit of $|S_{12}|$ with eqn. (1). The *inset* shows the calculated values of $f_1$ as a function of λ of the superconductor; the star shows the point corresponding to the experimental resonant frequency. (c) Experimental and calculated values of $f_1 - f_5$ as a function of mode number. The electric field distribution on the microstrip are shown in the *insets* for n=1 (upper left) and n=4 (lower right). (d) $|S_{12}|$ as a function of frequency around the fundamental resonance peak, $f_1$ as a function of temperature; the *inset* shows the variation of $f_1$ with temperature. (e) Temperature variation of $1/\lambda^2$ for the 60 nm NbN film; the solid line is a fit to the data with BCS temperature variation given by eqn. (2). The *inset* shows the temperature variation of $\lambda$. (f) Temperature variation of $1/\lambda^2$ obtained from NbN microstrip resonators made from films of different thickness; the zero-temperature value, $\lambda(0)$ obtained from the BCS fits (solid lines) are written next to each curve. (g) $|S_{12}|$ as a function of frequency around the fundamental resonance peak, $f_1$ as a function of temperature for the resonator made of the 90 nm thick $Nb_3Sn$; the *inset* shows the variation of $f_1$ with temperature. (h) Temperature variation of $1/\lambda^2$ obtained from the $Nb_3Sn$ microstrip resonator; the *inset* shows the temperature variation of $\lambda$.

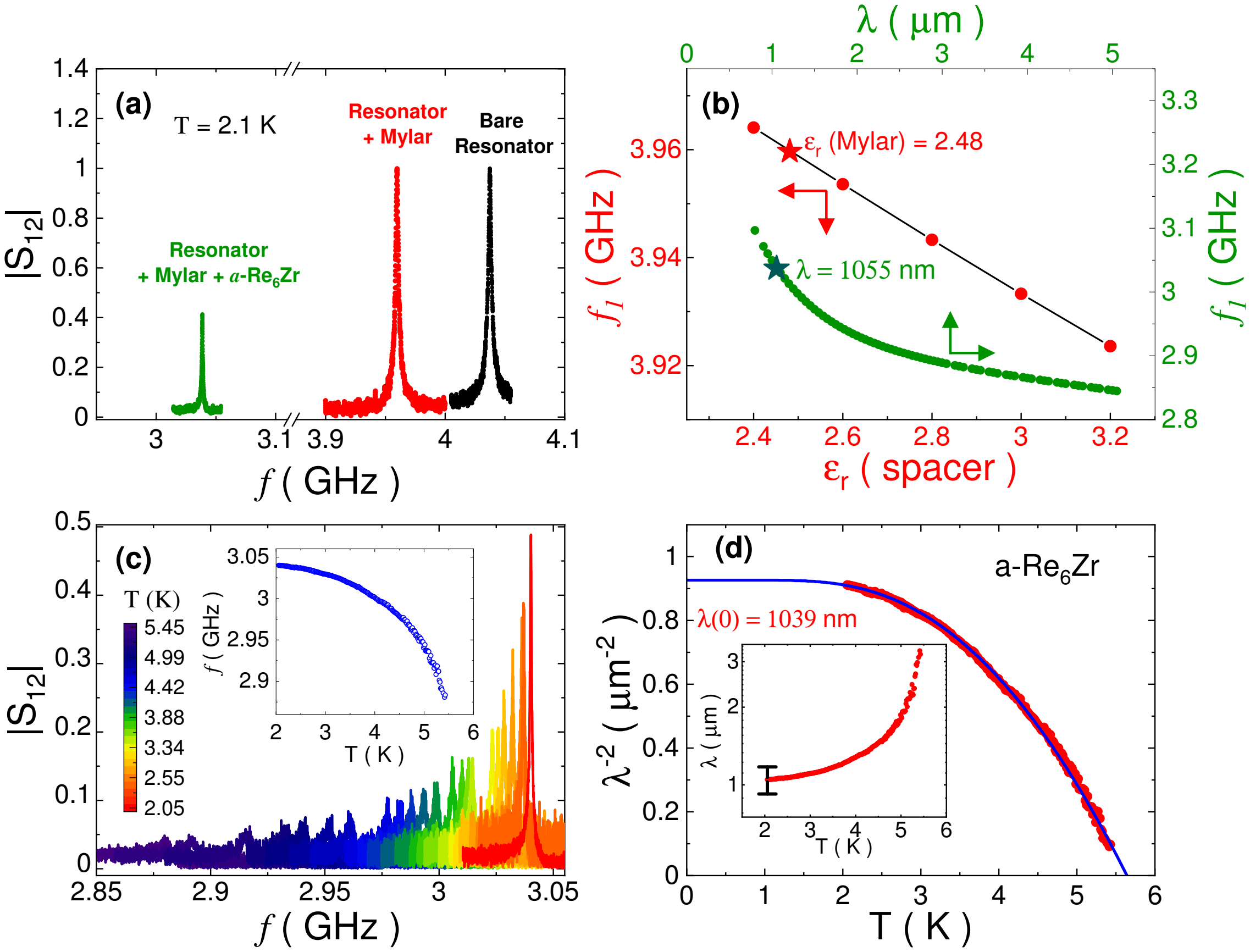

**Figure 3.** (a) Measured transmission coefficient $|S_{12}|$ at 2.1 K around the fundamental resonance peak in three configurations: (i) a bare NbN (thickness: 60 nm) microstrip resonator (black), (ii) with a 25 μm Mylar spacer layer placed on the microstrip (red), and (iii) with a 28 nm thick *a*-$Re_6Zr$ film placed upside-down on the Mylar sheet in flip-film geometry. (b) Calculated resonant frequency in configuration (ii) as a function of $\varepsilon_r$ of the spacer layer (red), and in configuration (iii) as a function of λ of *a*-$Re_6Zr$. The red and green stars are the points that correspond to the experimental resonant frequencies in configurations (ii) and (iii) respectively. (c) $|S_{12}|$ at different temperatures for the fundamental resonance peak in configuration (iii); the *inset* shows the temperature variation of $f_1$. (d) Temperature variation of $1/\lambda^2$ of the a-$Re_6Zr$ thin film along with the fit to the BCS temperature variation (solid line); the inset shows the temperature variation of $\lambda$.

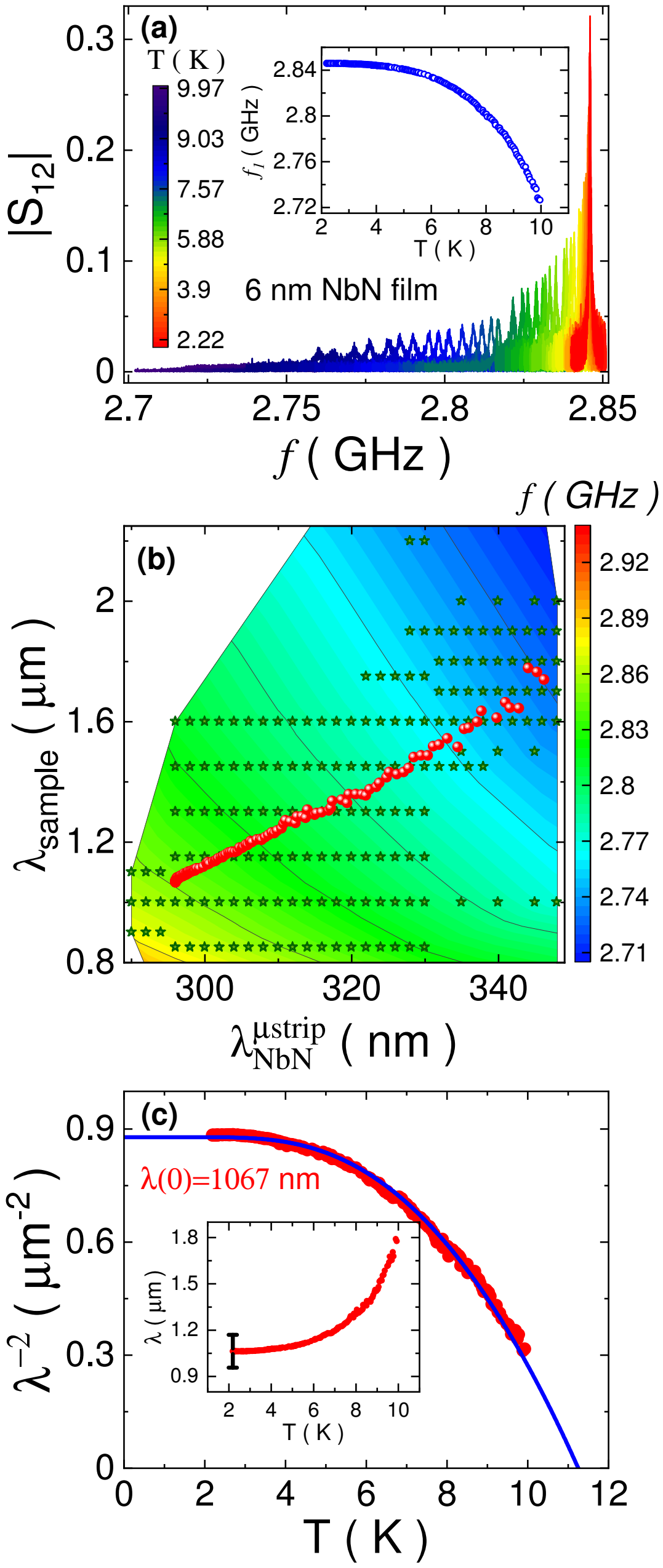

**Figure 4.** (a) Measured transmission coefficient $|S_{12}|$ around the fundamental resonance peak of a 60 nm thick NbN microstrip with a 6 nm thick NbN film placed upside-down with a 25 μm Mylar spacer in flip-film geometry. The *inset* shows the temperature variation of $f_1$. (b) Computed $f_1$ for this configuration shown as 3D intensity plot for different values of $\lambda_{NbN}^{\mu strip}$ and $\lambda_{sample}$; the stars correspond to points for which $f_1$ was calculated and the smooth color scale is a 2D interpolation of $f_1$ based on these points. The red circles correspond to the values that correspond to the measured values of $f_1$ at different temperatures. (c) Temperature variation of $1/\lambda^2$ of the 6 nm NbN thin film along with the fit to the BCS temperature variation (solid line); the inset shows the temperature variation of $\lambda$.

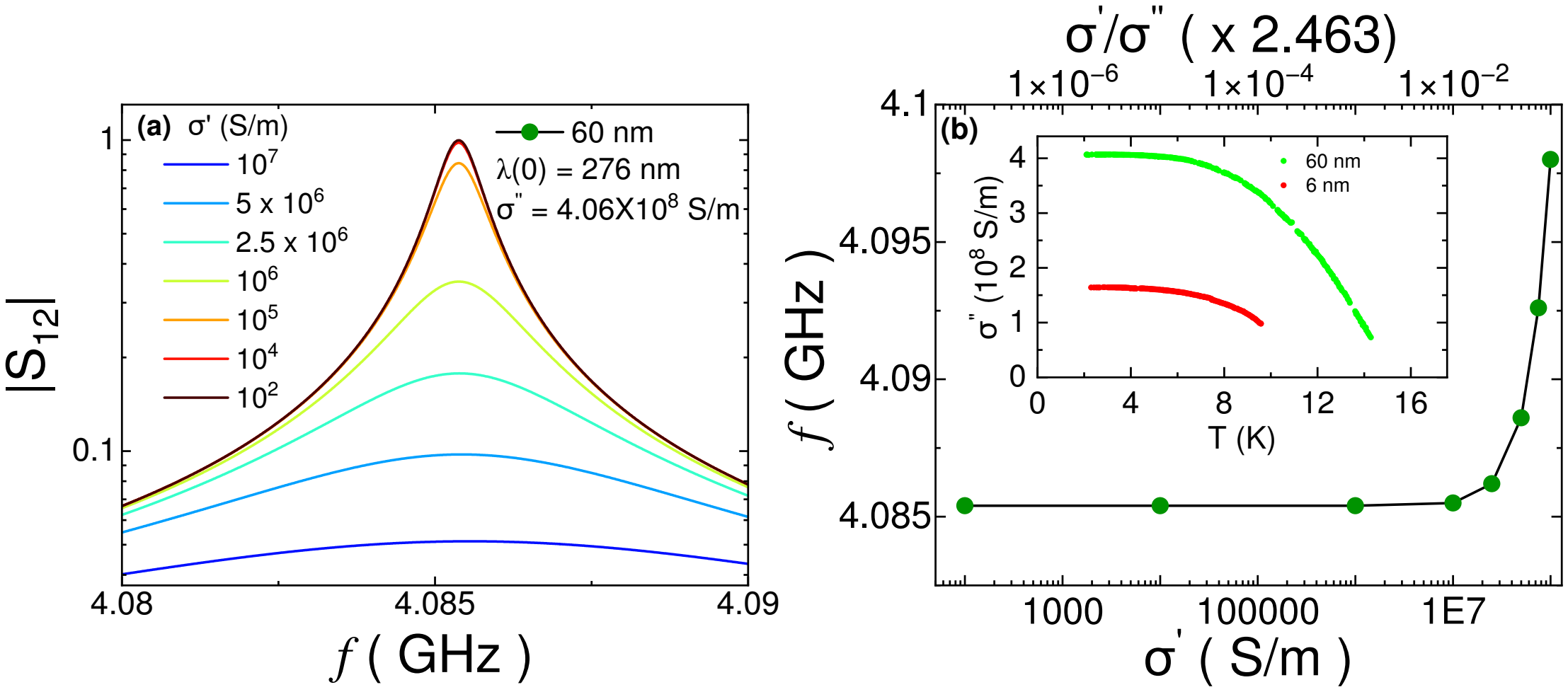

**Figure 5.** (a) Simulated transmission coefficient $|S_{12}|$ as a function of frequency for the fundamental frequency of a bare microstrip resonator for different values of $\sigma'$. $\sigma''$ is fixed to the value corresponding to the 60 nm thick NbN microstrip resonator at the lowest temperature. (b) Variation of $f_1$ as a function of $\sigma'$ (bottom axis) and $\frac{\sigma'}{\sigma''}$ (top axis). The inset shows the temperature variation of $\sigma''$ for the 60 nm and the 6 nm thick NbN microstrip resonators.

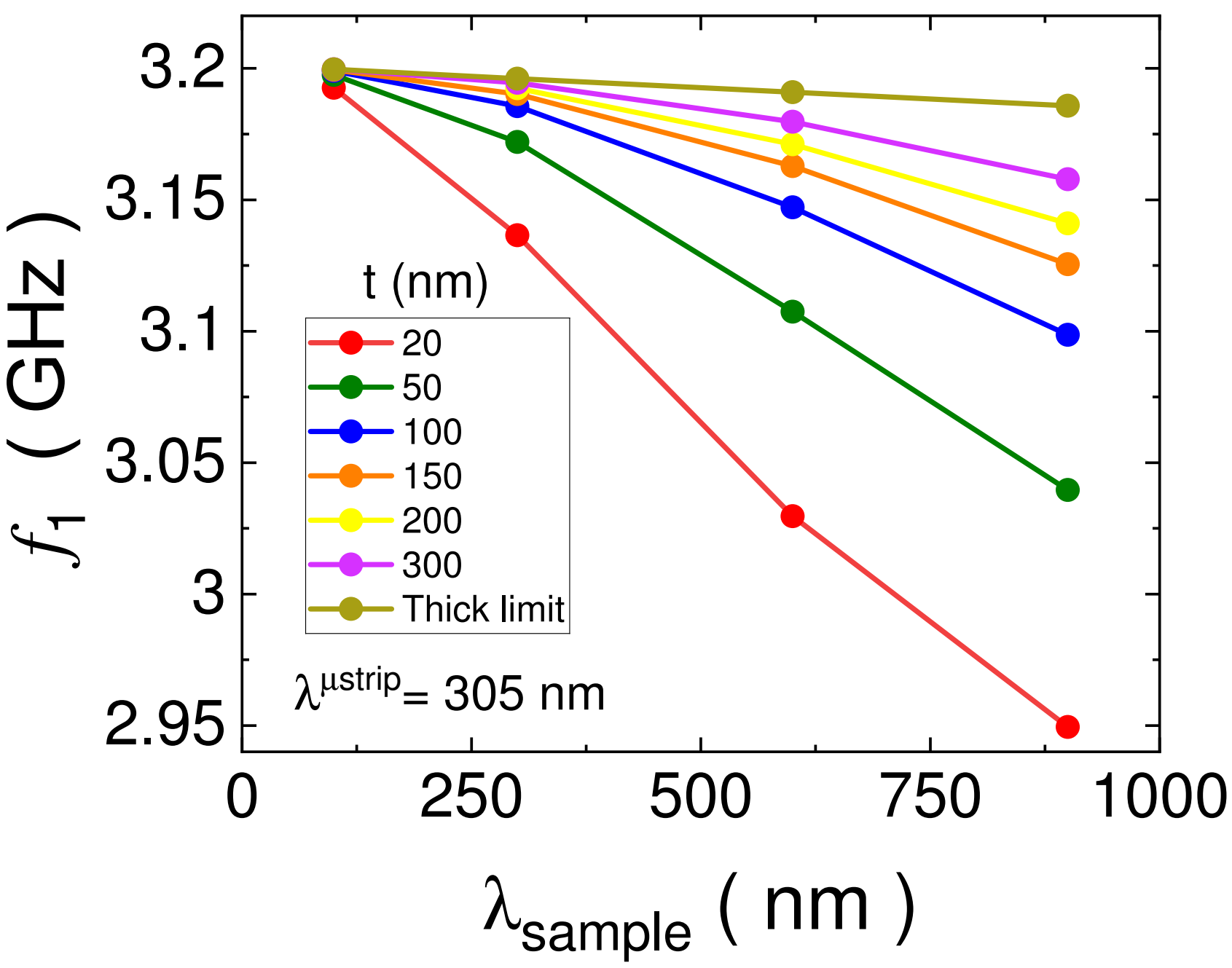

**Figure 6.** Calculated $f_1$ as a function of $\lambda_{\text{sample}}$ for flip-film geometry. Different curves correspond to different sample thickness, t. The penetration depth and thickness of the microstrip are assumed to be 305 nm and 57 nm respectively.